\documentclass[useAMS,usenatbib,usegraphicx]{mn2e}

\usepackage{aas_macros,url,amssymb}

\title[Will Comet 209P/LINEAR Generate the Next Meteor Storm?]{Will Comet 209P/LINEAR Generate the Next Meteor Storm?}
\author[Quanzhi Ye and Paul A. Wiegert]{Quanzhi Ye\thanks{E-mail:
qye22@uwo.ca} and Paul A. Wiegert\\
Department of Physics and Astronomy, The University of Western Ontario,
London, Ontario, N6A 3K7 Canada}
\begin{document}

\date{Accepted 1969 December 31. Received 1969 December 31; in original form 1969 December 31}

\pagerange{\pageref{firstpage}--\pageref{lastpage}} \pubyear{2013}

\maketitle

\label{firstpage}

\begin{abstract}
Previous studies have suggested that Comet 209P/LINEAR may produce strong meteor activity on Earth on 2014 May 24; however, exact timing and activity level is difficult to estimate due to the limited physical observations of the comet. Here we reanalyze the optical observations of 209P/LINEAR obtained during its 2009 apparition. We find that the comet is relatively depleted in dust production, with $Af\rho$ at 1 cm level within eight months around its perihelion. This feature suggested that this comet may be currently transitioning from typical comet to a dormant comet. Syndyne simulation shows that the optical cometary tail is dominated by larger particles with $\beta \sim 0.003$. Numerical simulations of the cometary dust trails confirm the arrival of particles on 2014 May 24 from some of the 1798--1979 trails. The nominal radiant is at RA $122^{\circ} \pm 1^{\circ}$, Dec $79^{\circ} \pm 1^{\circ}$ (J2000) in the constellation of Camelopardalis. Given that the comet is found to be depleted in dust 
production, we concluded that a meteor storm (ZHR$\geq1000$) may be unlikely. However, our simulation also shows that the size distribution of the arrived particles is skewed strongly to larger particles. Coupling with the result of syndyne simulation, we think that the event, if detectable, may be dominated by bright meteors. We encourage observers to monitor the expected meteor event as it will provide us with rare direct information on the dynamical history of 209P/LINEAR which is otherwise irretrievably lost.
\end{abstract}


\section{Introduction}

Comet 209P/LINEAR was discovered on 2004 February 3 by Lincoln Near-Earth Asteroid Research (LINEAR) as an asteroidal object; its cometary nature was later noted on 2004 March 30 \citep{mcn04}. \citet[][p.129 and p.689]{jen06} first pointed out the possibility of meteor activity originating from this comet in the near future, predicting that the dust trails
produced by the comet may come within 0.0002 AU from the Earth on 2014 May 24 around 7h UT. More recent examinations by J\'{e}r\'{e}mie Vaubaillon supported this possibility, commenting that a Zenith Hourly Rate \citep[i.e. the number of meteors that an average observer would see in one hour, given that the sky is clear and dark, and the radiant is at the zenith, c.f.][]{kos90} of a few hundreds is likely due to the close
encounter of materials released by the comet from all its apparitions between 1803 and 1924, but a meteor storm (ZHR$\geq1000$) might also be possible\footnote{\url{http://www.imcce.fr/langues/en/ephemerides/phenomenes/meteor/DATABASE/209_LINEAR/2014/index.php}, retrieved 2013 Oct. 1.}. Alternative prediction by Mikhail Maslov\footnote{\url{http://feraj.narod.ru/Radiants/Predictions/209p-ids2014eng.html}, retrieved 2013 Oct. 1.} suggested an encounter of the 1763--1783 and 1898--1919 materials, with maximum ZHR to be approximately 100, but also indicated that ``(possibilities of) storm levels are far from being excluded''.

Estimating the exact timing and particularly the level of this event is difficult, due in part to the limited reported photometric measurements of Comet LINEAR itself. Here, we aim at verifying and refining the prediction of this event. This will be done by reanalyzing the optical data obtained during the 2009 apparition of 209P/LINEAR to examine the dust production activity of the comet. The result will then be used as a constraint to refine the numerical simulation for the meteor event.

\section{Observations}

The observational data come from two sources: (i) survey images obtained by the Catalina Sky Survey (CSS); and (ii) the images taken by Michael J\"{a}ger near Tivoli, Namibia.

The CSS data are obtained by the Catalina 0.68-m Schmidt (located near Tucson, AZ) and the Uppsala 0.5-m Schmidt (located at Siding Spring, Australia) and will be used to constrain the dust production rate. The CSS uses identical single unfiltered $4096\times4096$ CCDs for both telescopes with pixel size of 2.5'' for the Catalina Schmidt and 1.8'' for the Uppsala Schmidt. The exposure times are variable between 20 and 30 s. Although unfiltered observations are sometimes discouraged for comet photometry due to potential contamination of Swan band emissions from the comet, 209P/LINEAR does not show signs of being an active comet, and we are therefore convinced that its Swan emissions will not be strong enough to affect our photometric result. The data on December
2008 and May 2009 are reduced with the CMC-14 \citep{eva02}, the other two sets of data are reduced with the APASS photometric catalog comes with the UCAC 4 \citep{zac13}. The system error for the two catalogs are estimated to be better than $\sim 0.2$ mag.

The images from M. J\"{a}ger will be used to probe the distribution of particles at different sizes, because of a favorable viewing geometry at the time of observation which will allow the separation of different sizes on the images. A total of 19 frames were taken by a 0.14-m Astrograph with SXVF-H9 CCD (pixel size at 3.2''), with 130 s exposure of each frame. The best responsive wavelength of the CCD is between $\sim$400--750 nm. The images are combined into one ``master'' frame by taking the median and are astrometrically calibrated with UCAC 4 \citep{zac13}.

\section{Dust Production Rate}

The dust activity of a comet can be determined by the product of its albedo ($A$), filling factor of grains within the aperture ($f$), and linear radius of the aperture at the comet ($\rho$) \citep{ahe84}:

\begin{equation}
\label{afpeq}
Af\rho = \frac{4r^2\Delta^2}{\rho} \frac{F_{C}}{F_{\odot}}
\end{equation}

where $r$ is the heliocentric distance of the comet in AU, $\Delta$ is the geocentric distance of the comet (in the same unit of $\rho$, typically in km or cm), and $F_{C}$ and $F_{\odot}$ are the fluxes of the comet within the field of view as observed by the observer and the Sun at a distance of 1 AU. The photometric aperture size, or $2\rho/\Delta$, is determined by the threshold value that the flux reaches an asymptote.

The resulting measurements are summarized in Table~\ref{obs}. We do not see a clear $Af\rho$ variation with respect to heliocentric distance of the comet due to the very weak dust production from the comet which is close to our detection limit. As contrast, typical comets have $Af\rho$ around 1--100 m \citep[e.g.][Fig. 5]{ahe95}, which is more than two order of magnitude larger. From these data we conclude that the $Af\rho$ of 209P/LINEAR stays at 1 cm level throughout its perihelion passage.

\begin{table*}
\begin{center}
\caption{$Af\rho$ measurements with the CSS data. The $Af\rho$ errors are the sum of raw magnitude errors computed from the SNR and the estimated systematic errors (0.2 mag) of the catalog used for calibration. \label{obs}}
\begin{tabular}{llccccc}
\hline
Date (UT) & Observing by & $r_h$ & $\Delta$ & $\rho$ & $Af\rho$ \\
          &              & (AU) & (AU) & (km) & (cm) \\
\hline
2008 Dec. 22.43 & Catalina & 1.765 & 0.996 & 21671 & $0.8\pm0.3$ \\
2009 May 28.76 & Siding Spring & 1.106 & 0.445 & 9295 & $1.3\pm0.4$ \\
2009 Jul. 8.80 & Siding Spring & 1.473 & 0.639 & 11679 & $1.1\pm0.4$ \\
2009 Aug. 12.62 & Siding Spring & 1.804 & 0.900 & 11749 & $1.5\pm0.6$ \\
\hline
\end{tabular}
\end{center}
\end{table*}

The number of particles ejected by the comet, $Q_g$, in the radii bin of [$a_1$,$a_2$], can be expressed as a function of heliocentric distance $r_h$ \citep{vau05}. Following the discussion above, we eliminate the distance term and rewrite the equation in a numerically simplified form:

\begin{equation}
\label{qg2}
Q_g(a_1,a_2)=\frac{655 A_1(a_1,a_2) Af\rho}{8 \pi A_B j(\phi)[A_3(a_1,a_2)+1000A_{3.5}(a_1,a_2)]}
\end{equation}

where $A_x=(a_2^{x-s}-a_1^{x-s})/(x-s)$ for $x\neq s$ and $A_x=\ln(a_2/a_1)$ for $x=s$, with $s$ to be the size population index, $A_B$ is the Bond albedo and $j(\phi)$ is the normalized phase function. By using $Af\rho=1$ cm following our earlier analysis, $s=2.6$ following the value found for 1P/Halley by \citet{ful20} over the simulated size range from $10^{-5}$ to $10^{-1}$~m in radius, and $A_B=0.05$, we found $Q_g=2.2\times10^6$ particle/s.

\section{Dust Tail Modeling}

The dust particles released from a small body (i.e. where the gravity of body is negligible) are driven by the radiation pressure and the gravity of the Sun. The ratio of these two quantities is usually defined as $\beta$, which is inversely proportional to the product of particle density and size, or $\beta \propto (\rho r)^{-1}$. Since these particles continue to follow a Keplerian trajectory around the Sun, we can simulate the motion of a large number of particles at different $\beta$ and release times, and produce what is called a syndyne-synchrone diagram \citep[e.g.][]{fin68}.

\begin{figure*}
\includegraphics[width=\textwidth]{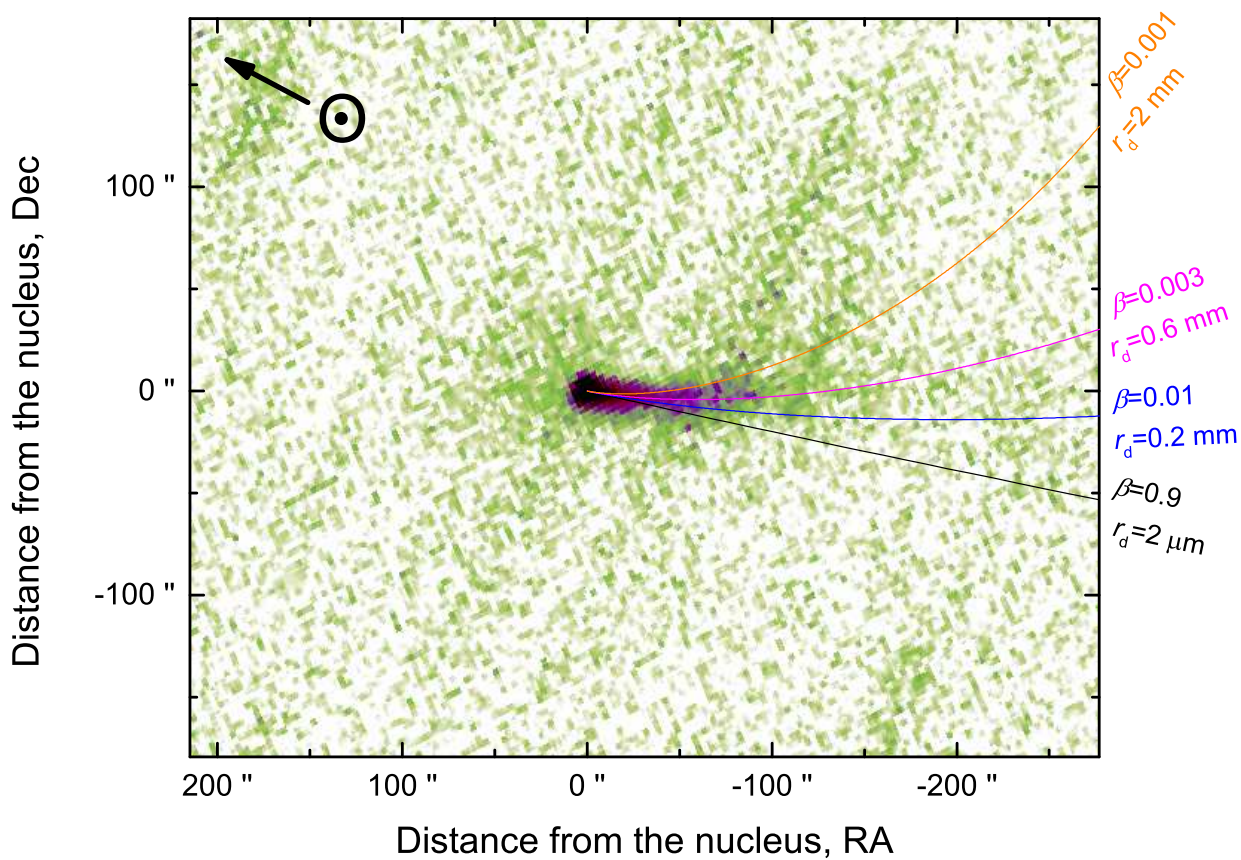}
\caption{209P/LINEAR on 2009 Apr. 25.10 (UT) as combined from 19 frames, with syndyne curves overimposed. The image is 4$\times$ magnified. The particle density is assumed at $\rho=300~\mathrm{kg \cdot m^{-3}}$.}
\label{fig-syndyne}
\end{figure*}

We compute the syndyne curves for the 2009 April 25 image and overimpose the modeling result to the image (Figure~\ref{fig-syndyne}). It can be seen that the particles dominate the optical tail have $\beta \sim 0.003$. Assuming a meteoroid density of $\rho=300$ kg$\cdot$m$^{-3}$ and a particle radius $r$ measured in meters, the ratio of solar radiation pressure to gravity $\beta$ is given by $\beta = 5.74 \times 10^{-4}/{\rho r}$ in these units following \citet{fox82,wil83}. This yields a particle size of $\sim 0.6$~mm, which indicates a predominance of large particles. This phenomenon is not attributed by scattering enhancement considering the Sun-Observer-Comet angle to be $\phi=98^{\circ}$ at the time of observation. Additional synchrone simulations show that the optical tail, measured $\sim 2'$ in length, was composed by particles released within the 15 days prior to the time of observation. The particle distribution seems to terminate at $\beta \sim 0.01$, indicating the dust size distribution does not 
follow a power law beyond this value. Such upper limit is quite low comparing to dynamically new comets, and is comparable to veteran comets such as 2P/Encke, 10P/Tempel and 67P/Churyumov-Gerasimenko which are also characterized by a smaller $\beta$ range with size distribution following the power law \citep{ful04}.

On the image, we also notice a fan-like brightness enhancement in the anti-solar direction. Possibilities such as processing artifacts or ghost images of nearby stars have been ruled out. Such enhancement cannot be explained by simple syndyne-synchrone model which assumes a null ejection velocity. We then examine M. J\"{a}ger's images and observations by other observers in the adjacent dates for a confirmation. These other images were not used for modeling work due to shorter accumulated exposure time (less than 1/3 of the one on Apr. 25). Unfortunately, the images on Apr. 22 were out of focus, and the comet was close to a bright star on Apr. 26. No similar feature was visible on the images on Apr. 28. However, on Seiichi Yoshida's comet observation collection\footnote{\url{http://www.aerith.net/comet/catalog/0209P/2009.html}, retrieved on 2013 Sep. 15.}, we do notice a possible brightness enhancement of $\Delta \mathrm{mag} \sim 1$ mag that occurred shortly before May 1, which is marginally higher than 
systematic fluctuation ($\lesssim 0.5$ mag as inferred from the chart). On the other hand, the IAU Minor Planet Center collected no observations from Apr. 23 to May 14\footnote{\url{http://www.minorplanetcenter.net/db_search/show_object?object_id=209P}, retrieved 2013 Sep. 15.}. We suspect that this feature, if indeed physically real, may indicates an ejection event, but conclusion cannot be drawn with the absence of additional observation.

\section{Dynamical Modeling of Meteoroid Streams}

The details of dynamical modeling of this work are similar to those given in \citet{wie13} and \citet{ye13}. Here we only summarize the key concepts.

The simulation is conducted using the RADAU method \citep{eve85} with a time step of seven days. We include the eight major planets, with the Earth-Moon system represented by a single particle at the barycenter of the two bodies. The initial conditions of the eight planets were derived from the JPL DE405 ephemeris \citep{sta98}. We first integrate the orbit of 209P/LINEAR backwards 250 years. The comet is then integrated forward again, with particles released at each perihelion passage. The number of particles is assumed to follow the different size distribution $dN/dr \propto r^{-2.6}$ as our earlier calculation for $Q_g$. Post-Newtonian general relativistic corrections and radiative (i.e. Poynting-Robertson) effects are also included. The ratio of solar radiation pressure to gravitational force, $\beta$, is related to the particle radius $r$ (in $\mu$m) through $\beta = 1.9/r$, \citep{weijac93}, assuming a particle mass density $\rho = 300~\mathrm{kg \cdot m^{-3}}$. All meteoroids which have a
close encounter of $\lesssim 0.02$ AU from the Earth are collected. Then, the forward integration is repeated, with particles released only near the initial conditions for the meteoroid collected in the first step. The second-generation particles are given a random change of up to 10\% in each velocity component, and those passed closest to the Earth in space and time will be considered to contribute to the simulated outburst.

We consider the comet to be active when its heliocentric distance is less than 2.3~AU where water ice sublimation is expected to start \citep[c.f.][]{mcn04}. The particles are released following the cometary ejection model described by \citet{jon95}. We use an absolute total magnitude $M_1=16.7$ and absolute nuclear magnitude $M_2=19.8$ for 209P/LINEAR from the JPL Small-Body Database\footnote{\url{http://ssd.jpl.nasa.gov/sbdb.cgi?sstr=209P}, retrieved 2013 Aug. 24.}, which are derived from 546 observations assuming a magnitude slope of 14 and 5 respectively. The water production rate, $\log{Q_{\mathrm{H_2 O}}}=26.73 \pm 0.52$, can be calculated by the formula proposed by \citet{jor92}. In the absence of other details, we assume a Bond albedo for the nucleus of 0.05, which yields a nucleus size of 600 m; we also assume a nucleus density of $300~\mathrm{kg \cdot m^{-3}}$.

In our simulation, the amount of water sublimation is assumed to be directly proportional to the amount of solar heating the nucleus absorbs and the gas-to-dust ratio is taken to be unity. This yields a total of $9 \times 10^9$~kg of gas per perihelion passage for this comet. The mean production rate is 300~kg/s or $10^{28}$~molec/s, which varies by a factor of about $(2.3/0.87)^2 \approx 7$ between perihelion (0.87~AU) and the start of gas production (2.3~AU). An equal amount of dust is assumed to be released between the sizes of $10^{-9}$ to $10^{-1}$~m in radius, though only particles larger than $10^{-5}$~m are simulated: given our assumed size distribution, this gives a fraction $>0.99999$ of the mass and $4 \times 10^{-7}$ of the number of particles simulated. Our mean simulated gas production rate is thus about $20\times$ times that proposed by \citet{jor92} for a comet of this magnitude. Coupled with the low $Af\rho$ values measured, we expect that our simulations will overestimate the likely 
meteoroid production of this comet though it must be recognized that the earlier perihelion passages which produce particles intersecting the Earth in 2014 may have benefited from higher gas production rates than the comet currently displays.

\section{Discussion}

The reanalysis of optical observations of 209P/LINEAR in its 2009 apparition shows that the comet is largely inactive. We examine the reported $Af\rho$ values of other short-period comets with $q \lesssim 1.3$~AU, which reveals that 209P/LINEAR is among the ones with lowest perihelion $Af\rho$. Interestingly, some comets at similar $M_1$ or nucleus diameter have perihelion $Af\rho$ more than one order of magnitude larger\footnote{$Af\rho$ measurements are collected from CometasObs, \url{http://www.astrosurf.com/cometas-obs/}, retrieved 2013 Oct. 1.}, such as 26P/Grigg-Skjellerup ($M_1=16.7$) and 76P/West-Kohoutek-Ikemura \citep[$D=0.66$~km as given by][]{lam04}. This indicates that 209P/LINEAR may be a transitional object from typical comet and a dormant comet, or its nucleus is significantly smaller than we thought.

\begin{figure*}
\includegraphics[width=\textwidth]{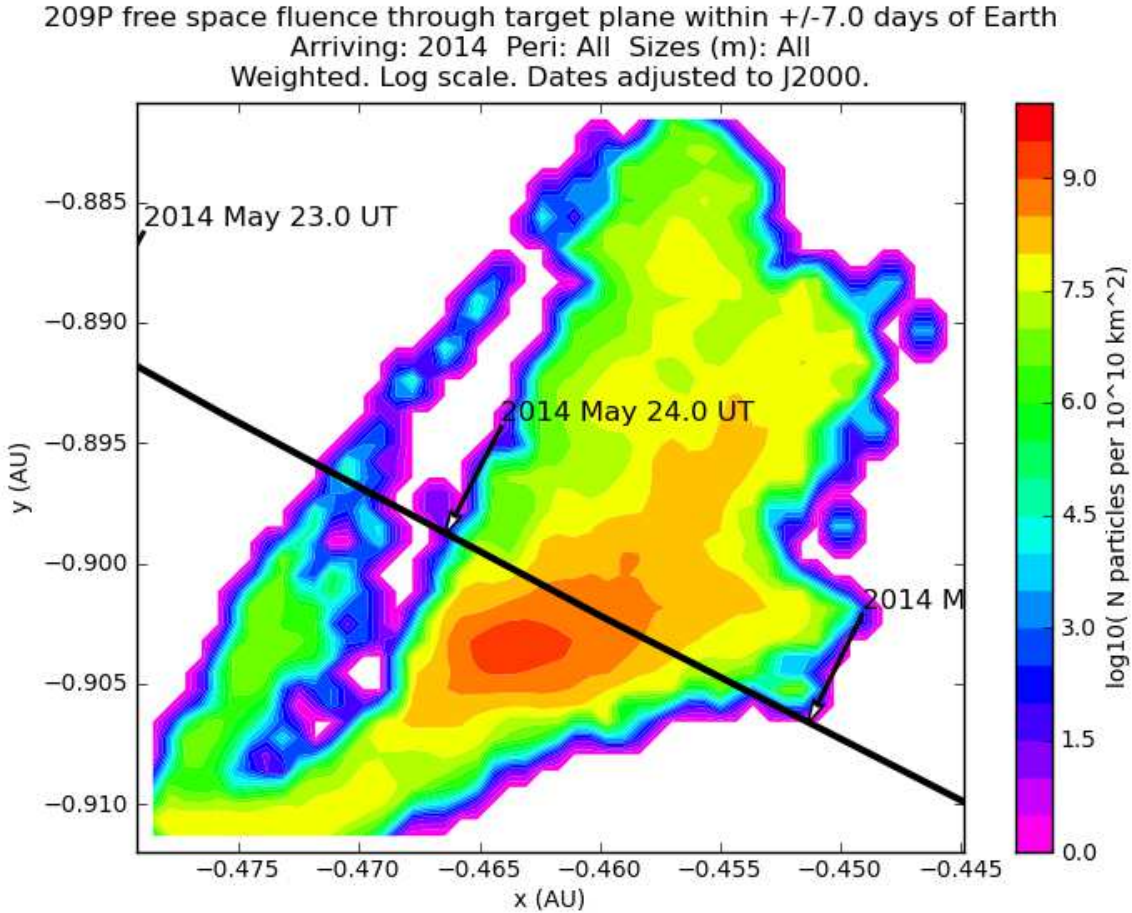}
\caption{The footprint of the meteoroid stream from 209P/LINEAR projected on the ecliptic. The colour scheme labels the free space (no gravitational focusing) fluence of particles through a plane perpendicular to the stream's arrival direction. Locations of
the Earth at particular times are labeled with arrows. The Sun is to the upper right.}
\label{fig-footprint}
\end{figure*}

Our simulations of the comet trails confirm the arrival of a number of trails from the parent on the 2014 May 24. The nodal footprint of the stream is shown in Figure~\ref{fig-footprint}. Particles arriving were produced on perihelion passages in 1798, 1803, 1868, 1878, 1883 as well as those occuring from 1924 to 1954 and 1964 to 1979. This differs in some detail from earlier reports (Vaubaillon and Maslov) but is perhaps not unexpected. Our simulations show a relatively close approach between 209P/LINEAR and Jupiter occurred in 1976--1977 which shifted the comet's perihelion outwards by 0.1~AU. Such close encounters make predictions beyond them into the past much more difficult, so some differences between our simulations and those of others are to be expected. Despite the differing details, we will see that our results are consistent in an overall sense with those of earlier investigators.

\begin{figure}
\includegraphics[width=0.5\textwidth]{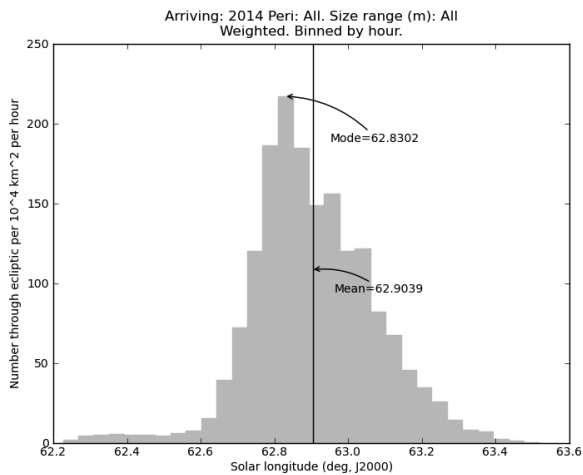}
\caption{The expected hourly rates of meteor arrivals at the Earth per $10^4$~km$^2$. The peak is expected to occur at a solar longitude of 62.83$^{\circ}$ (appropriate to 2004 May 24.27 UT, or 6h29m) with a Full-Wide-Half-Maximum (FWHM) of about 0.4 days.}
\label{fig-sollon}
\end{figure}

The number of particles arriving at the Earth during the shower is shown in Figure~\ref{fig-sollon}. The number is shown per hour per $10^4$~km$^2$ of collecting area which approximates ZHR that might be expected for a visual observer under the radiant (which we expect at high northerly latitudes, RA $122^{\circ} \pm 1^{\circ}$, Dec $79^{\circ} \pm 1^{\circ}$ in J2000, i.e. in the constellation of Camelopardalis) with a clear sky approximately 100~km on a side available for viewing. From this graph we derive a predicted ZHR of about 200 for our nominal scenario. However, given the current relatively weak dust production of the comet, rates could be much lower. Of course, absent information on the activity of the parent body during the perihelion passages during which the arriving meteoroids were produced we cannot say much about expected rates, nevertheless we conclude that a meteor storm is unlikely. We do encourage observers to monitor the expected shower as it will provide us with rare direct information 
on the dust activity of the parent in the past, information which is otherwise irretrievably lost.

\begin{figure}
\includegraphics[width=0.5\textwidth]{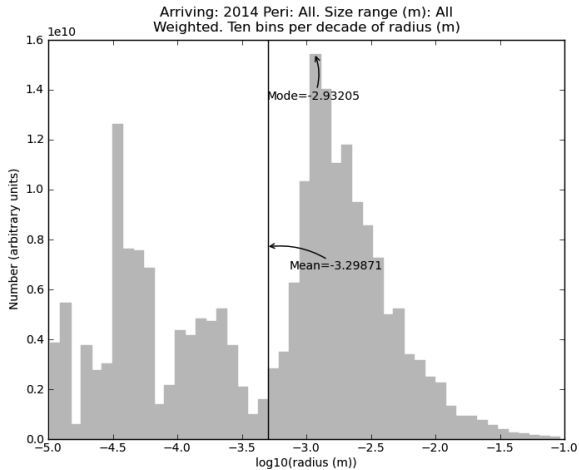}
\caption{The expected size distribution.}
\label{fig-sizes}
\end{figure}

Though the number of particles arriving at Earth is relatively small, the size distribution is skewed strongly towards larger particles (Figure~\ref{fig-sizes}). Despite the relative rarity of large particles in the simulation (due to a size distribution which favors smaller ones), meteoroids arriving at Earth are predominately larger than 1~mm. Given that our syndyne calculations indicate that the parent produces particles most abundantly at the size which are most efficiently delivered to Earth in our simulations, it may be that the shower will prove unusual for the number of bright meteors produced.

\section{Summary}

We reanalyzed the optical observations made during the 2009 apparition of 209P/LINEAR to constraint the dust production activity of the comet, in the hope to verify and refine the prediction of the forthcoming meteor outburst in 2014 as caused by this comet. Our analysis showed that 209P/LINEAR is considerably depleted in dust production, with $Af\rho \approx 1$~cm within eight months around its perihelion, which indicated the comet may be currently transitioning from typical comet to a dormant comet. By fitting the observation to syndyne model, we found that the tail is dominated by larger particles. The upper limit of a power-law distribution of $\beta$ is found to be $\sim0.01$.

Our numerical simulation confirmed the arrival of particles from some of the 1798--1979 cometary trails from 209P/LINEAR on 2014 May 24. The peak of the meteor activity is expected at 2014 May 24, 6h29m UT, with FWHM about 0.4 days. The meteor rate is very difficult to estimate due to our poor knowledge of the comet's physical property and dynamical history in particular, but given that the comet is relatively depleted in dust production, we concluded that a meteor storm may be unlikely. However, our simulation showed that the size selection is skewed strongly to larger particles; considering that the syndyne simulation indicated that the tail of 209P/LINEAR is dominated by larger particles, we suggested that the meteor outburst, if detectable, may be dominated by bright meteors. Observations of the outburst will give us crucial information about the dynamical past of 209P/LINEAR which is otherwise irretrievably lost.

\section*{Acknowledgments}

We thank Iwan Williams and an anonymous referee for their comments. We also thank Eric Christensen, who kindly made the CSS data available to us, as well as Andrea Boattini and Robert McNaught for their services on data acquisition. The CSS survey is funded by the National Aeronautics and Space Administration under Grant No. NNH12ZDA001N-NEOO, issued through the Science MIssion Directorate's Near Earth Object Observations Program. We also thank Michael J\"{a}ger for allowing us to use his images. We gratefully acknowledge Peter Jenniskens, Esko Lyytinen, J\'{e}r\'{e}mie Vaubaillon and Mikhail Maslov for their exploratory works. Q.Y. thanks Peter Brown, his supervisor, for supporting him to explore various scientific interests. He also thanks Man-To Hui for helps and discussions regarding the syndyne-synchrone models. This work was performed in part with the support of the Natural Sciences and Engineering Research Council of Canada.

\bibliographystyle{mn2e}
\bibliography{man}

\begin{thebibliography}{}

\bibitem[\protect\citeauthoryear{{A'Hearn}, {Millis}, {Schleicher}, {Osip} \&
  {Birch}}{{A'Hearn} et~al.}{1995}]{ahe95}
{A'Hearn} M.~F.,  {Millis} R.~L.,  {Schleicher} D.~G.,  {Osip} D.~J.,
  {Birch} P.~V.,  1995, \icarus, 118, 223

\bibitem[\protect\citeauthoryear{{A'Hearn}, {Schleicher}, {Millis}, {Feldman}
  \& {Thompson}}{{A'Hearn} et~al.}{1984}]{ahe84}
{A'Hearn} M.~F.,  {Schleicher} D.~G.,  {Millis} R.~L.,  {Feldman} P.~D.,
  {Thompson} D.~T.,  1984, \aj, 89, 579

\bibitem[\protect\citeauthoryear{{Evans}, {Irwin} \& {Helmer}}{{Evans}
  et~al.}{2002}]{eva02}
{Evans} D.~W.,  {Irwin} M.~J.,    {Helmer} L.,  2002, \aap, 395, 347

\bibitem[\protect\citeauthoryear{{Everhart}}{{Everhart}}{1985}]{eve85}
{Everhart} E.,  1985, in {Carusi} A.,  {Valsecchi} G.~B.,  eds, Dynamics of
  Comets: Their Origin and Evolution {An efficient integrator that uses
  Gauss-Radau spacings}.
Kluwer, Dordrecht, pp 185--202

\bibitem[\protect\citeauthoryear{{Finson} \& {Probstein}}{{Finson} \&
  {Probstein}}{1968}]{fin68}
{Finson} M.~J.,  {Probstein} R.~F.,  1968, \apj, 154, 327

\bibitem[\protect\citeauthoryear{{Fox}, {Williams} \& {Hughes}}{{Fox}
  et~al.}{1982}]{fox82}
{Fox} K.,  {Williams} I.~P.,    {Hughes} D.~W.,  1982, \mnras, 200, 313

\bibitem[\protect\citeauthoryear{{Fulle}}{{Fulle}}{2004}]{ful04}
{Fulle} M.,  2004, {Motion of cometary dust}.
pp 565--575

\bibitem[\protect\citeauthoryear{{Fulle}, {Levasseur-Regourd}, {McBride} \&
  {Hadamcik}}{{Fulle} et~al.}{2000}]{ful20}
{Fulle} M.,  {Levasseur-Regourd} A.~C.,  {McBride} N.,    {Hadamcik} E.,  2000,
  \aj, 119, 1968

\bibitem[\protect\citeauthoryear{{Jenniskens}}{{Jenniskens}}{2006}]{jen06}
{Jenniskens} P.,  2006, {Meteor Showers and their Parent Comets}

\bibitem[\protect\citeauthoryear{{Jones}}{{Jones}}{1995}]{jon95}
{Jones} J.,  1995, 275, 773

\bibitem[\protect\citeauthoryear{{Jorda}, {Crovisier} \& {Green}}{{Jorda}
  et~al.}{1992}]{jor92}
{Jorda} L.,  {Crovisier} J.,    {Green} D.~W.~E.,  1992, in AAS/Division for
  Planetary Sciences Meeting Abstracts \#24 Vol.~24 of Bulletin of the American
  Astronomical Society, {The Correlation Between Cometary Water Production
  Rates and Visual Magnitudes}.
p.~1006

\bibitem[\protect\citeauthoryear{{Koschack} \& {Rendtel}}{{Koschack} \&
  {Rendtel}}{1990}]{kos90}
{Koschack} R.,  {Rendtel} J.,  1990, WGN, Journal of the International Meteor
  Organization, 18, 44

\bibitem[\protect\citeauthoryear{{Lamy}, {Toth}, {Fernandez} \&
  {Weaver}}{{Lamy} et~al.}{2004}]{lam04}
{Lamy} P.~L.,  {Toth} I.,  {Fernandez} Y.~R.,    {Weaver} H.~A.,  2004, {The
  sizes, shapes, albedos, and colors of cometary nuclei}.
pp 223--264

\bibitem[\protect\citeauthoryear{{McNaught} \& {Kocer}}{{McNaught} \&
  {Kocer}}{2004}]{mcn04}
{McNaught} R.~H.,  {Kocer} M.,  2004, \iaucirc, 8314, 1

\bibitem[\protect\citeauthoryear{Standish}{Standish}{1998}]{sta98}
Standish E.~M.,  1998, Technical report, Planetary and Lunar Ephemerides
  {DE405/LE405}.
NASA Jet Propulsion Laboratory

\bibitem[\protect\citeauthoryear{{Vaubaillon}, {Colas} \& {Jorda}}{{Vaubaillon}
  et~al.}{2005}]{vau05}
{Vaubaillon} J.,  {Colas} F.,    {Jorda} L.,  2005, \aap, 439, 751

\bibitem[\protect\citeauthoryear{{Weidenschilling} \&
  {Jackson}}{{Weidenschilling} \& {Jackson}}{1993}]{weijac93}
{Weidenschilling} S.~J.,  {Jackson} A.~A.,  1993, Icarus, 104, 244

\bibitem[\protect\citeauthoryear{{Wiegert}, {Brown}, {Weryk} \&
  {Wong}}{{Wiegert} et~al.}{2013}]{wie13}
{Wiegert} P.~A.,  {Brown} P.~G.,  {Weryk} R.~J.,    {Wong} D.~K.,  2013, \aj,
  145, 70

\bibitem[\protect\citeauthoryear{{Williams} \& {Fox}}{{Williams} \&
  {Fox}}{1983}]{wil83}
{Williams} I.~P.,  {Fox} K.,  1983, in {Lagerkvist} C.-I.,  {Rickman} H.,  eds,
  Asteroids, Comets, and Meteors {The evolution of meteor streams}.
pp 399--409

\bibitem[\protect\citeauthoryear{Ye, Wiegert, Brown, Campbell-Brown \&
  Weryk}{Ye et~al.}{2013}]{ye13}
Ye Q.~Z.,  Wiegert P.~A.,  Brown P.~G.,  Campbell-Brown M.,    Weryk R.~J.,
  2013, \mnras, in review

\bibitem[\protect\citeauthoryear{{Zacharias}, {Finch}, {Girard}, {Henden},
  {Bartlett}, {Monet} \& {Zacharias}}{{Zacharias} et~al.}{2013}]{zac13}
{Zacharias} N.,  {Finch} C.~T.,  {Girard} T.~M.,  {Henden} A.,  {Bartlett}
  J.~L.,  {Monet} D.~G.,    {Zacharias} M.~I.,  2013, \aj, 145, 44

\end{thebibliography}

\label{lastpage}

\end{document}